\documentclass[12pt]{article}

\usepackage[margin=1in]{geometry}
\usepackage{setspace}

\usepackage{amsmath, amssymb, mathtools}
\usepackage{amsthm}

\usepackage{microtype}
\usepackage{mathptmx}
\usepackage[american]{babel}
\usepackage{csquotes}
\usepackage[style=apa,backend=biber,natbib=true]{biblatex}
\DeclareLanguageMapping{american}{american-apa}
\usepackage{enumitem}

\usepackage{placeins}
\usepackage{graphicx}
\usepackage{float}
\usepackage{tabularx}
\usepackage{longtable}
\usepackage{booktabs}

\usepackage[hidelinks]{hyperref}

\theoremstyle{definition}
\newtheorem{theorem}{Theorem}[section]
\newtheorem{proposition}[theorem]{Proposition}
\newtheorem{corollary}[theorem]{Corollary}
\newtheorem{definition}[theorem]{Definition}
\newtheorem{postulate}[theorem]{Postulate}
\newtheorem{assumption}[theorem]{Assumption}
\newtheorem{axiom}[theorem]{Axiom}
\newtheorem{lemma}[theorem]{Lemma}
\newtheorem{remark}[theorem]{Remark}

\newtheorem{convention}[theorem]{Convention}

\title{The Impossibility of Cohesion Without Fragmentation}
\author{Daisuke Hirota\\ Independent Researcher}
\date{}

\begin{document}

\maketitle

\begin{abstract}
Accounts of social division and cohesion presuppose relations whose
maintenance already depends on positional compatibility. We develop this
feasibility layer as a static theory: a positioning event fixes players'
positions, and a compatibility function determines which pairs can
sustain a relation. Two results follow from one formalism. First, any
non-uniform compatibility pattern on the dyads sharing a positional
requirement forces fragmentation and cohesion together; the only events
that guarantee no collapse are those rendering every pair compatible.
The asymmetry between unilateral severance and bilateral confirmation
recovers the convention underlying pairwise stability without a behavioral premise, while a
separate specialization, in which compatibility partitions the position
space, yields assortative survival without choice or opportunity.
Second, at the measure level, the survival rate factorizes
exactly into integration and cohesion terms. Along any complete
relaxation path that relaxes one pair at a time (the generic case under
distance thresholds with absolutely continuous positions), the cohesion
term is non-monotone once at least three players are present, and in
clustered configurations it is bounded above by a sharp asymptotic value
that finite systems approach but never attain, with strict crossing of
the two terms once the number of clusters exceeds two.
\end{abstract}


\section{Introduction}
\label{sec:introduction}

\subsection{The Phenomenon}
\label{subsec:phenomenon}

Why do division and cohesion in social systems recurrently arise together?
As political blocs polarize, their internal alignment sharpens; as
religious communities differentiate from surrounding society, mutual
ties tighten among their members; as firms bifurcate into competing
technical standards, internal coordination on each side becomes
tighter. These pairings appear across disparate empirical settings,
yet a first-principles account of \emph{why they must arise as a
pair} is lacking.

The question is structural: what feature of relation formation makes
the co-emergence of division and cohesion mechanically necessary,
prior to any specific behavioral, affective, or strategic dynamic
operating over the relations?

\subsection{Misattribution and the Feasibility Layer}
\label{subsec:misattribution}

Existing paradigms of relation formation locate the mechanism
downstream of relational feasibility. In network-formation games
\citep{jackson1996strategic}, links form as functions of utility,
presupposing that a connection is possible whenever mutually
preferred. In probabilistic models
\citep{watts1998collective, barabasi1999emergence}, links emerge with
a specified probability, subsuming the conditions of infeasibility
into stochastic noise. In behavioral and dynamical accounts
\citep{schelling1971dynamic, cartwright1956structural}, links evolve
over time according to tolerance thresholds or the resolution of
imbalanced cycles, presupposing the ties whose sign or presence is
to be adjusted.

Each of these paradigms treats a substantive question about the
relation (strength, probability, sign, or dynamic sign-flip) while
implicitly answering a prior question in the affirmative: whether
the relation is structurally possible. When positions in a social
space are governed by a compatibility function, some pairs are
structurally incompatible, and the maintenance of the relation is
impossible regardless of preference, probability, or strategy. We
call the space of these prior structural possibilities the
\emph{feasibility layer}: the layer within which any downstream
dynamic on relations must proceed. When an outcome is already forced
at this layer, attributing it to a downstream
mechanism---preference, probability, or strategy---is a
misattribution: the mechanism selects among the outcomes the layer
admits, but does not produce the necessity itself.
Section~\ref{sec:related} positions the theory relative to specific
accounts.

\subsection{Contributions}
\label{subsec:contributions}

This paper develops the feasibility layer in three parts.

\medskip
\noindent\textbf{(i) Reformulation as feasibility.}
Section~\ref{sec:model} formalizes relation maintenance as a problem
of positional compatibility: the maintenance of a relation between
players $i$ and $j$ depends on $f_g(E(i), E(j))$, evaluated at the
positions fixed by a positioning event $E$. Positional compatibility
is the structural filter conditioning any downstream dynamic on
relations.

\medskip
\noindent\textbf{(ii) Existence-level results (Section~\ref{sec:existence}).}
Whenever the realized compatibility pattern is non-uniform on the
shared-relevant dyads (non-degeneracy;
Definition~\ref{def:g-relevant-degeneracy}), fragmentation and
occurrence of local cohesion\footnote{``Cohesion'' is used at three
levels---dyad-axis, existence, and network density;
Remark~\ref{rem:three-tier-cohesion} calibrates them.}
arise as complementary outcomes of the
same compatibility function (Theorem~\ref{thm:relativity}); the
impossibility of the title is conditional in exactly this sense. The
duality is asymmetric
in logical form: collapse is triggered by a disjunction of
withdrawal conditions, occurrence of local cohesion by a
conjunction (Proposition~\ref{prop:veto-asymmetry}). This asymmetry
recovers the unilateral-severance/bilateral-confirmation convention
of strategic network formation without a behavioral premise
(Remark~\ref{rem:pairwise-stability}); independently, the
equivalence-relation specialization yields assortative survival from
maintenance feasibility alone, with neither a choice nor a
contact-opportunity component (Remark~\ref{rem:homophily}). The only
positioning events
guaranteeing no $g$-induced relational collapse for every admissible
minimum-condition profile and initial relation set are the
universally compatible ones---those
under which every player pair is $g$-compatible
(Theorem~\ref{thm:impossibility}); and any confirmed shared
condition necessitates at least one severed pair
(Corollary~\ref{cor:outgroup-inescapability}).

\medskip
\noindent\textbf{(iii) Measure-level results (Section~\ref{sec:measure}).}
The survival rate factorizes exactly into integration and cohesion:
$\sigma = I \cdot K$ (Proposition~\ref{prop:decomposition}). Along
every complete strict relaxation path (from a single edge to the
complete graph, with every transition simple), cohesion is
necessarily non-monotone for $N \ge 3$
(Theorem~\ref{thm:universal-non-monotone}); in $(s, S, D)$-clustered
configurations of $k \ge 2$ equal-size clusters, the crossing of
integration and cohesion is quantitatively controlled with $K(D)$
bounded by the sharp asymptotic value $(3k - 2)/k^2$ (strict
crossing for $k \ge 3$; non-strict at $k = 2$;
Theorem~\ref{thm:structural-crossing}); and monotone
cohesion trajectories are characterized exactly by a synchronization
condition on the transitions
(Theorem~\ref{thm:synchronisation}).

\section{Related Work and Positioning}
\label{sec:related}

Mathematical sociology has long distinguished formal structural
constraints from the behavioral dynamics they enable
\citep{skvoretz2000looking}. The present theory operates on the
structural side of this distinction, on a layer that most existing
accounts of division and cohesion implicitly presuppose.

\subsection{Dynamical and Behavioral Genealogies}
\label{subsec:dynamical}

\paragraph{Segregation dynamics.}
Schelling's segregation model derives macro-level separation from
individual tolerance thresholds applied to local neighborhood
composition \citep{schelling1971dynamic}. The present theory yields
separation with no threshold and no behavioral parameter: once positions
are fixed by a positioning event, the division into compatible
classes---an exact partition under the equivalence-relation
specialization (Proposition~\ref{prop:exclusive-regions})---is
determined by the compatibility function alone. Schelling gives a
behavioral generative mechanism for segregation; the framework here
fixes the structural boundary within which any such mechanism must
operate.

\paragraph{Structural balance.}
Structural balance theory characterizes when signed networks admit
factional partitions in the absence of imbalanced cycles
\citep{cartwright1956structural, aref2019balance}. Dynamical extensions
describe how such partitions emerge as termini of imbalance-reducing
processes \citep{antal2005dynamics, marvel2011continuous}. Both
traditions presuppose the existence of signed ties; the present theory
operates one level earlier, on whether a tie is feasible at all. The
mutually exclusive regions of
Proposition~\ref{prop:exclusive-regions}, with the inter-class
severance established in Section~\ref{sec:existence}, induce---under
the canonical signed completion of Remark~\ref{rem:davis-factions}
(positive within, negative between classes)---the same vertex
partition that balance theory identifies as factions, obtained here
statically and without a signed-relation primitive.
Recent extensions of the balance-dynamics program derive a critical
level of interconnectedness above which societies fragment into
internally cohesive, mutually hostile sub-communities
\citep{pham2020effect}; such phase-transition results operate above the
feasibility layer analyzed here, and their trajectories are governed
by the compatibility partition delineated in
Section~\ref{sec:existence}.

\paragraph{Partisan sorting and tie dissolution.}
At the empirical level, the alignment of social identities with
partisanship \citep{mason2018uncivil} and the associated dissolution of
cross-cutting ties following partisan position-taking
\citep{levendusky2009partisan, klar2016independent} instantiate the
structural pattern formalized here: position fixation under a
non-uniform realized compatibility pattern coincides with selective
collapse at the feasibility layer, prior to any claim that identity
alignment causes animosity.

\paragraph{Affective polarization.} A recent literature models the emergence of cohesion,
polarization, and fragmentation as dynamical or affective processes
\citep{iyengar2019affective, nettasinghe2025outgroup}. These models
operate on the feasibility layer characterized here: any dynamic of
opinion, emotion, or strategic choice unfolds over relations whose
maintenance is already governed by positional compatibility.
Corollary~\ref{cor:outgroup-inescapability} isolates the prior
fact that whenever the realized compatibility pattern is
non-uniform on the shared-relevant dyads, a positioning event
confirms local cohesion only by severing at least one \emph{incompatible dyad}
(a pair $\{i,j\}$ with $f_g(E(i),E(j)) = 0$); under the
equivalence-relation specialization
(Corollary~\ref{cor:outgroup-partition-equivalence-case}), this
incompatibility structure sharpens into a genuine
in-group/out-group partition on which the affective literature
operates.

\paragraph{Non-monotone polarization.}
A model of content filtering on social networks derives non-monotonic
comparative statics in network density, with sharp transitions between
moderation and polarization \citep{campbell2025polarization}. There the
non-monotonicity issues from a behavioral mechanism---selective
forwarding along ideological lines---operating over existing ties;
Theorem~\ref{thm:universal-non-monotone} locates a structurally
parallel non-monotonicity in the maintained-relation density itself,
prior to any such mechanism.

\subsection{Adjacent Formalisms}
\label{subsec:feasibility}

The measurement of cohesion and community structure in networks has
its own tradition, distinct from the feasibility question addressed
here: centrality-based cohesion measures~\citep{freeman1978centrality},
group-membership formalizations~\citep{freeman1992sociological},
node-connectivity structural cohesion~\citep{moody2003structural}, and
the wider community-detection literature~\citep{fortunato2010community}
all take the presence of ties as given and quantify or partition
them. The intra-component density $K = M/Q$ used in
Section~\ref{sec:measure} is one such measure; its role in this
paper is as the pair-density factor in the $\sigma = I \cdot K$
decomposition, whose non-monotonicity along positional relaxations
is the object of study.

\paragraph{Compatibility graphs and constraint satisfaction.}
\phantomsection\label{subsec:csp-positioning}
Each active dyad-axis pair defines a binary constraint on player
positions in the sense of binary constraint networks
\citep{montanari1974networks,dechter2003constraint}; a positioning
event is an assignment, and the maintained-relation network
records which relational constraints are satisfied by that
assignment. Our contribution lies one layer above the
constraint-satisfaction apparatus itself. We take that apparatus as
given and identify its social-theoretic consequences under partial
satisfaction of a fixed assignment: the existence-level duality of
complementary necessity and the measure-level decomposition
$\sigma = I \cdot K$ under threshold-defined filtrations. Latent-space network models
\citep{hoff2002latent} and random geometric graph filtrations
\citep{penrose2003random} share graph-theoretic scaffolding but
address different questions.

\section{Model}
\label{sec:model}

\subsection{Design Principles}
\label{subsec:design-principles}

\begin{postulate}[Relational layer fixation]
\label{po:layer-fixation}
The analysis is conducted relative to a fixed relational layer,
defined by a specific set of gain axes $G$ and minimum condition
function $a$. The initial relation set
$\mathcal{R}_{\textnormal{initial}}$ contains exactly the relations
belonging to this layer. Two consequences of layer fixation are
adopted as modeling assumptions: within a layer, a player's minimum
conditions are partner-independent ($a$ is a function of the player
and the axis alone), and at most one relation is associated with
each unordered player pair. Partner-specific requirement profiles
are represented as distinct layers.
\end{postulate}

\begin{postulate}[Reduction to pairwise relations]
\label{po:reduct-pairwise}
The theory represents all social relations as pairwise relations:
every relation is dyadic, and the set of potential dyads is
$\binom{\mathcal{P}}{2}$. An initial relational layer may contain
relations on any subset of these dyads.
\end{postulate}

The analysis proceeds through logical (not temporal) comparison in
two modes. Its basic form (Section~\ref{sec:existence})
compares pre- and post-event states with fixed compatibility. Its
parametric extension (Section~\ref{sec:measure},
via~\S\ref{subsec:relaxation}) is a monotone family of such
comparisons parameterized by the strictness of the compatibility
function; parameter values do not carry temporal ordering. The
structural limits identified here apply to relations that carry
content, are embedded in a social category, or are maintained within
a community, all of which operate under position-dependent conditions.

\subsection{Structural Primitives}
\label{subsec:primitives}

\begin{definition}[Primitives]
\label{def:player-set}
Let $\mathcal{P}$ denote the finite universal set of players, with
$N := |\mathcal{P}| \ge 2$, and let $G$ denote the finite set of
gain axes, given as axiomatic input; each gain axis represents a
type of value or condition realizable through a relation
(recognition, belonging, monetary transfer, authority, etc.). The
local player set of a pairwise relation $R$ is a dyad
$P_R \in \binom{\mathcal{P}}{2}$, typically $\{i,j\}$.
\end{definition}

\begin{definition}[Minimum condition function]
$a : \mathcal{P} \times G \to \{0,1\}$, fixed at the level of the
relational layer (Postulate~\ref{po:layer-fixation}).
\end{definition}

\begin{axiom}[Minimum condition]
\label{ax:minimum-condition}
For any $i \in \mathcal{P}$ and $g \in G$: $a(i,g) = 1$ if and only
if player $i$ withdraws from the relation whenever $g$ is absent.
\end{axiom}

\begin{remark}[Structural interpretation of withdrawal]
\label{rem:withdrawal}
Withdrawal is a structural predicate: configurations in which $g$
fails are excluded from the domain of viable relations for player
$i$. It is not a behavioral or psychological notion.
\end{remark}

\begin{definition}[Active gain axes]
\label{def:active-axes}
For a relation $R$ with player set $P_R$,
$G_{\textnormal{active}}(R) := \{g \in G \mid \exists i \in P_R : a(i,g)=1\}$.
We consider only relations with $|G_{\textnormal{active}}| > 0$.
\end{definition}

\subsection{Position Space and Compatibility}
\label{subsec:position-compatibility}

\begin{definition}[Position space]
Let $\mathcal{L}$ denote the set of positions a player may occupy,
common to all gain axes; the position of player $i$ is
$\ell_i \in \mathcal{L}$. Positions represent physical location,
opinion, role, or affiliation; no topological, metric, or ordering
structure is imposed unless explicitly stated.
\end{definition}

\begin{definition}[Compatibility function]
\label{def:compatibility-function}
\label{po:symmetry}
For a gain axis $g$, the compatibility function is
$f_g : \mathcal{L} \times \mathcal{L} \to \{0, 1\}$, required to be
symmetric: $f_g(\ell, \ell') = f_g(\ell', \ell)$ for all
$\ell, \ell' \in \mathcal{L}$. A gain axis $g$
is satisfied between $i$ and $j$ at positions $\ell_i, \ell_j$ iff
$f_g(\ell_i, \ell_j) = 1$.
\end{definition}

\begin{remark}[Constraint-network reading]
\label{rem:csp-reading}
In constraint-network terms (Section~\ref{subsec:csp-positioning}):
positions are values, a positioning event is an assignment, each
$f_g$ is a symmetric binary constraint, and
Definition~\ref{def:axis-satisfaction} records which constraints
the assignment satisfies.
\end{remark}

\subsection{Dependency Types and Equivalence Structure}
\label{subsec:dependency-types}

\begin{assumption}[Non-Null Compatibility]
\label{ass:non-null}
Every $g$ appearing in any relation in
$\mathcal{R}_{\textnormal{initial}}$ satisfies
$\exists\,(\ell, \ell') \in \mathcal{L}^2
    : f_g(\ell, \ell') = 1$.
\end{assumption}

\begin{definition}[Dependency types of gain axes]
\label{def:dependency-classification}
The dependency type of $g$ is:
\begin{itemize}
    \item \textbf{Existence-dependent}: $f_g \equiv 1$ on
    $\mathcal{L} \times \mathcal{L}$.
    \item \textbf{Position-dependent}: $f_g \not\equiv 1$ and
    $\exists\,(\ell,\ell') : f_g(\ell,\ell') = 1$.
\end{itemize}
\end{definition}

\begin{proposition}[Position-dependent gain axes and exclusive regions]
\label{prop:exclusive-regions}
Let $g$ be a position-dependent gain axis. If $f_g$ defines an
equivalence relation on $\mathcal{L}$, then $\mathcal{L}$ partitions
into mutually exclusive regions $\{L_1, L_2, \dots, L_k\}$ with
$k \ge 2$, such that
\[
\ell_i \in L_a,\; \ell_j \in L_b,\; a \neq b
    \;\Longrightarrow\; f_g(\ell_i, \ell_j) = 0.
\]
\end{proposition}

\begin{proof}
The equivalence classes of $f_g$ on $\mathcal{L}$ partition
$\mathcal{L}$ into the exclusive regions. The condition
$f_g \not\equiv 1$ (position-dependence) ensures at least two
classes, so $k \ge 2$. For $\ell_i \in L_a$ and $\ell_j \in L_b$
with $a \neq b$, the pair $(\ell_i, \ell_j)$ is not equivalent under
$f_g$, so $f_g(\ell_i, \ell_j) = 0$.
\end{proof}

\begin{remark}[Homophily without a choice component]
\label{rem:homophily}
Homophily---the tendency for similar players to remain
connected---is conventionally decomposed into \emph{choice}
homophily, driven by a preference for similar others, and
\emph{induced} homophily, driven by the opportunity structure
\citep{mcpherson2001birds}. The present framework identifies a third
formal source of assortative survival: \emph{maintenance
feasibility}. When the compatibility function $f_g$ induces an
equivalence relation on the position space
(Proposition~\ref{prop:exclusive-regions}), only within-class
relations survive a positioning event, while cross-class relations
sever. Cross-class relations may already be present in the initial
layer, so the resulting sorting is generated neither by a preference
for similar others nor by differential contact opportunity; it
arises because some existing relations are structurally
non-maintainable once positions are fixed. Whether maintenance
feasibility is interpreted empirically as an institutional,
technological, legal, or opportunity constraint lies outside the
formal result.
\end{remark}

\subsection{Positioning Events and Maintenance}
\label{subsec:positioning-networks}

\begin{convention}[Edge-set typing]
\label{conv:edge-set-typing}
Every edge-set on $\mathcal{P}$ ($\mathcal{E}$, $\mathcal{E}^{-}$,
$\mathcal{E}^{+}$, $\mathcal{E}_g^{-}$, $\mathcal{E}_t$) is a subset
of $\binom{\mathcal{P}}{2}$, the set of unordered distinct pairs.
When a formula writes $f_g(E(i), E(j))$ for a pair $\{i,j\}$, the
notation denotes the symmetric evaluation of $f_g$ on the two
positions.
\end{convention}

\begin{definition}[Relations within a layer]
\label{def:complete-relation}
Within the fixed relational layer
$(G, a, \mathcal{L}, \{f_g\})$ (Postulate~\ref{po:layer-fixation}),
a relation $R$ is identified by its dyad
$P_R \in \binom{\mathcal{P}}{2}$; we write $R_{ij}$ for the
relation on $\{i,j\}$ when it exists, and use the accessors
$G_R := G$ and $a_R := a|_{P_R \times G}$.
\end{definition}

\begin{definition}[Positioning event]
\label{def:positioning-event}
\label{ax:unique-position}
A \emph{positioning event} is a map $E : \mathcal{P} \to \mathcal{L}$
that forces each player to occupy a determinate position: after $E$,
$\ell_i = E(i)$ for all $i$. Single-valuedness is substantive, not
merely notational: a player cannot occupy two mutually incompatible
positions simultaneously within a structural snapshot (this
commitment is audited in Appendix~\ref{app:independence}).
\end{definition}

\begin{remark}[Scope and exogeneity]
\label{rem:scope-exogeneity}
$E$ is treated as an exogenous position-fixing operator. The theory
addresses the structural consequences that follow once positions are
fixed; the strategic, cultural, or stochastic mechanisms generating
$E$ are outside its scope.
\end{remark}

\begin{definition}[Initial relation set]
\label{def:initial-relation-set}
$\mathcal{R}_{\textnormal{initial}} := \{R_1, \dots, R_m\}$: the set
of relations existing prior to $E$. Throughout, an initial relation
set is called \emph{admissible} if it is nonempty and every relation
in it satisfies the standing requirements of
Assumption~\ref{ass:non-null} and
Definition~\ref{def:active-axes}.
\end{definition}

\begin{definition}[Axis-level satisfaction under $E$]
\label{def:axis-satisfaction}
Let $R = R_{ij}$ be the relation on $\{i,j\}$, and let $g
\in G$ be position-dependent. Write $R \models_E g$ if
\[
\bigl(a(i,g)=1 \lor a(j,g)=1\bigr)
    \;\Longrightarrow\; f_g(E(i),E(j)) = 1,
\]
and $R \not\models_E g$ otherwise. The symbol $\models_E$ is a
per-axis abbreviation; $R \models_E g$ holds vacuously when
$a(i,g) = a(j,g) = 0$.
\end{definition}

\begin{definition}[Post-event maintenance]
\label{def:post-event-maintenance}
Let $R = R_{ij}$ be the relation on $\{i,j\}$. $R$ is
\emph{maintained after $E$} iff $R \models_E g$ for every
position-dependent $g \in G$.
\end{definition}

\begin{definition}[Maintained relation set]
\label{def:maintained-relation-set}
\[
\mathcal{R}_{\textnormal{maintained}}(E) :=
    \{R \in \mathcal{R}_{\textnormal{initial}}
    \mid R \text{ is maintained under } E\}.
\]
\end{definition}

\begin{definition}[Confirmed shared condition]
\label{def:confirmed-shared-condition}
$g \in G$ is a \emph{confirmed shared condition} for $\{i,j\}$ under
$E$ if $a(i,g) = a(j,g) = 1$ and $f_g(E(i),E(j)) = 1$.
\end{definition}

\begin{remark}[Three levels of cohesion]
\label{rem:three-tier-cohesion}
``Cohesion'' operates at three levels, and the analogous
existence-level predicates are deliberately minimal. (i) The
dyad-axis predicate of a confirmed shared condition
(Definition~\ref{def:confirmed-shared-condition}) is strictly
stronger than $R \models_E g$, requiring $a(i,g)=a(j,g)=1$. (ii)
\emph{Local cohesion} (Definition~\ref{def:local-cohesion-occurrence}
below) is its existence predicate over the initial layer; a
confirmed shared condition does not by itself imply maintenance in
the multi-axis model, since the witness relation may collapse
through another active axis, but when $g$ is the witness's unique
active position-dependent axis---as throughout
Convention~\ref{conv:complete-dyad}---the witness is maintained and
local cohesion is carried by a surviving tie. (iii) The
network-level density $K$ of Definition~\ref{def:sigma-I-K} is
strictly stronger than the mere occurrence of local cohesion.
Symmetrically, fragmentation
(Definition~\ref{def:fragmentation} below) records the loss of at
least one relation and does not by itself imply disconnection or
macroscopic partition: the minimal-witness convention matches the
logical form of the impossibility results of
Section~\ref{sec:existence}, while network-level integration and
cohesion are measured in Section~\ref{sec:measure}.
\end{remark}

\begin{definition}[Initial edge set and graph invariants]
\label{def:graph-invariants}
The set of \emph{initial dyads} is
$\mathcal{E}^{-} := \{\{i,j\} \in \binom{\mathcal{P}}{2} \mid
R_{ij} \in \mathcal{R}_{\textnormal{initial}}\}$. For a graph
$\mathcal{N} = (\mathcal{P}, \mathcal{E})$ on the player set, with
$\mathrm{CC}(\mathcal{N})$ its set of connected components (maximal
mutually reachable vertex sets), define the edge count and the
intra-component pair count as
\[
M(\mathcal{N}) := |\mathcal{E}|,
\qquad
Q(\mathcal{N}) :=
    \sum_{C \in \mathrm{CC}(\mathcal{N})} \binom{|C|}{2}.
\]
Singletons contribute $\binom{1}{2} = 0$ to $Q$.
\end{definition}

\begin{definition}[Fragmentation]
\label{def:fragmentation}
\emph{Fragmentation} induced by $E$ occurs if $\exists\, R \in
\mathcal{R}_{\textnormal{initial}}$ with $R \notin
\mathcal{R}_{\textnormal{maintained}}(E)$.
\end{definition}

\begin{definition}[Axis-induced fragmentation]
\label{def:g-fragmentation}
Fix a position-dependent gain axis $g$. \emph{$g$-induced
fragmentation} $\operatorname{Frag}_g(E)$ occurs if
$\exists\, R \in \mathcal{R}_{\textnormal{initial}}$ with $g \in G_R$
and $R \not\models_E g$. $\operatorname{Frag}_g(E)$ implies
fragmentation of Definition~\ref{def:fragmentation}; the converse
fails, since global fragmentation may be witnessed by a different
axis $h \neq g$.
\end{definition}

\begin{definition}[Occurrence of local cohesion]
\label{def:local-cohesion-occurrence}
Let $g \in G$ be a gain axis. \emph{Local cohesion with respect to
$g$} induced by $E$ occurs if $\exists\, R \in
\mathcal{R}_{\textnormal{initial}}$ with $P_R = \{i,j\}$ such that
$g \in G_R$ and $g$ is a confirmed shared condition for $\{i,j\}$
under $E$. Unqualified, \emph{local cohesion} induced by $E$ occurs
if it occurs with respect to some $g \in G$.
\end{definition}

\subsection{Relaxation Family and Threshold Transitions}
\label{subsec:relaxation}

As announced in Section~\ref{subsec:design-principles},
Section~\ref{sec:measure} extends the fixed-$f_g$ comparison of
Section~\ref{sec:existence} to a one-parameter family indexed by
the strictness of the compatibility function: comparative statics,
with no temporal ordering among parameter values.

\begin{definition}[Relaxation family]
\label{def:relaxation-family}
A \emph{relaxation family} is a one-parameter family of compatibility
functions $\{f_t\}_{t \in T}$, $T \subseteq \mathbb{R}$, such that
$t \le t'$ implies
$f_t(\ell, \ell') \le f_{t'}(\ell, \ell')$
for all $\ell, \ell' \in \mathcal{L}$. Each $f_t$ is symmetric by
Definition~\ref{def:compatibility-function}. The parameter $t$ is the \emph{laxity}
of the constraint.
\end{definition}

\begin{definition}[Threshold graph]
\label{def:threshold-graph}
Given $E$, the threshold graph at parameter $t$ is
$\mathcal{G}_t := (\mathcal{P}, \mathcal{E}_t)$ with
\[
\mathcal{E}_t := \{\, \{i,j\} \in \binom{\mathcal{P}}{2}
    : f_t(E(i), E(j)) = 1 \,\}.
\]
\end{definition}

\begin{remark}[Nestedness]
\label{rem:nestedness}
By Definition~\ref{def:relaxation-family}, $t \le t'$ implies
$\mathcal{E}_t \subseteq \mathcal{E}_{t'}$; the threshold graphs
form a monotone chain of subgraphs of the complete graph on
$\mathcal{P}$.
\end{remark}

\begin{definition}[Threshold transition]
\label{def:threshold-transition}
Because $\mathcal{P}$ is finite and the edge sets are nested
(Remark~\ref{rem:nestedness}), the distinct realized edge sets form
a finite chain
$\mathcal{E}^{(0)} \subsetneq \mathcal{E}^{(1)} \subsetneq \cdots
    \subsetneq \mathcal{E}^{(m)}$.
A \emph{threshold transition} is the passage from
$\mathcal{E}^{(r-1)}$ to $\mathcal{E}^{(r)}$. Write $\mathcal{G}_r^-$
and $\mathcal{G}_r^+$ for the threshold graphs immediately before and
after the $r$-th transition, and set
$\Delta M_r := M(\mathcal{G}_r^+) - M(\mathcal{G}_r^-) \ge 1$ and
$\Delta Q_r := Q(\mathcal{G}_r^+) - Q(\mathcal{G}_r^-) \ge 0$
(Definition~\ref{def:graph-invariants}).
No assumption is made that transitions are attained at parameter
values.
\end{definition}

\begin{definition}[Metric specialization]
\label{def:metric-specialization}
When $(\mathcal{L}, \mathrm{dist})$ is a metric space, set
$f_d(\ell, \ell') := \mathbf{1}\{\mathrm{dist}(\ell, \ell') \le d\}$.
This family is symmetric automatically, and its transitions are
attained at the finitely many realized distances.
\end{definition}

\section{Existence-Level Results}
\label{sec:existence}

\subsection{Structural Necessity of Fragmentation and Cohesion}
\label{subsec:necessity}

\begin{definition}[$g$-relevant edges and degeneracy under $E$]
\label{def:g-relevant-degeneracy}
Let $\mathcal{R}_{\textnormal{initial}}$ be a set of dyadic relations
on $\mathcal{P}$. For a gain axis $g$, define the set of
$g$-shared-relevant initial edges by
\begin{multline*}
\mathcal{E}_g^{-} := \{\, \{i,j\}\in \binom{\mathcal{P}}{2}
    \mid \exists\, R \in \mathcal{R}_{\textnormal{initial}}
    \text{ with } P_R = \{i,j\}, \\
    a_R(i,g)=1 \land a_R(j,g)=1 \,\}.
\end{multline*}
We say that $g$ is \emph{degenerate under $E$} if the map
$\phi : \{i,j\}\mapsto f_g(E(i),E(j))$ is constant on
$\mathcal{E}_g^{-}$. Degeneracy is a property of the triple
$(g, E, \mathcal{R}_{\textnormal{initial}})$, since
$\mathcal{E}_g^{-}$ depends on the initial relation set; we say
``$g$ is non-degenerate under $E$'' when
$\mathcal{R}_{\textnormal{initial}}$ is fixed by context.
\end{definition}

\begin{theorem}[Complementary necessity of fragmentation and local cohesion]
\label{thm:relativity}
Let $\mathcal{R}_{\textnormal{initial}}$ be a nonempty set of
relations and $E: \mathcal{P} \to \mathcal{L}$ a positioning event.
Consider a position-dependent gain axis $g$ with $\mathcal{E}_g^{-}
\neq \emptyset$.
\begin{enumerate}
    \item \textbf{The degenerate case:} the compatibility map
    $\{i,j\} \mapsto f_g(E(i), E(j))$ is constant on $\mathcal{E}_g^{-}$.
    When the constant value is~$1$, $g$ forces no collapse among the
    $g$-shared-relevant relations (pairs in $\mathcal{E}_g^{-}$);
    when it is~$0$, all of them collapse uniformly.
    \item \textbf{The structural outcome (complementary necessity):}
    if $g$ is not degenerate under $E$, then $E$ necessarily produces
    both $g$-induced fragmentation and an occurrence of local
    cohesion with respect to $g$.
    \begin{itemize}
      \item \textbf{Fragmentation.} Every $R_{ij}$ collapses whenever
      $\{i,j\} \in \mathcal{E}_g^{-}$ and $f_g(E(i),E(j))=0$.
      \item \textbf{Occurrence of local cohesion.} Every relation
      $R_{uv} \in \mathcal{R}_{\textnormal{initial}}$ produces a
      confirmed shared condition for $g$ (Definition~\ref{def:confirmed-shared-condition})
      whenever $\{u,v\} \in \mathcal{E}_g^{-}$ and $f_g(E(u),E(v))=1$.
    \end{itemize}
\end{enumerate}
\end{theorem}

\begin{proof}
Assume $g$ is not degenerate under $E$. Then the image of $\phi$
restricted to $\mathcal{E}_g^{-}$ is $\{0, 1\}$, so there exist pairs
$\{i,j\}, \{u,v\} \in \mathcal{E}_g^{-}$ with $f_g(E(i),E(j))=0$ and
$f_g(E(u),E(v))=1$.

\emph{Fragmentation.} For such $\{i,j\}$, since $\{i,j\} \in
\mathcal{E}_g^{-}$, $a(i,g)=a(j,g)=1$, so the antecedent of
$R_{ij} \models_E g$ fires non-vacuously. But $f_g(E(i),E(j))=0$,
so $R_{ij} \not\models_E g$. By the single-valuedness of $E$
(Definition~\ref{ax:unique-position}),
positions are fixed, so this failure cannot be resolved: $R_{ij}
\notin \mathcal{R}_{\textnormal{maintained}}(E)$.

\emph{Occurrence of local cohesion.} For such $\{u,v\}$, there exists $R_{uv} \in
\mathcal{R}_{\textnormal{initial}}$ with $a(u,g)=a(v,g)=1$, and
$f_g(E(u), E(v))=1$. Both conditions of
Definition~\ref{def:confirmed-shared-condition} hold; $g$ is a
confirmed shared condition for $\{u,v\}$.
\end{proof}

Degeneracy is evaluated on $\mathcal{E}_g^{-}$, where the
requirement of $g$ is bilateral. A dyad on which exactly one player
requires $g$ lies outside $\mathcal{E}_g^{-}$ and, by
Definition~\ref{def:axis-satisfaction}, still collapses whenever
$f_g(E(i),E(j)) = 0$; Case~1 therefore controls the
$g$-shared-relevant relations only. This gap between the disjunctive
collapse condition and the conjunctive confirmation condition is
exactly Proposition~\ref{prop:veto-asymmetry} below.

\subsection{Asymmetry of Structural Veto}
\label{subsec:veto}

\begin{proposition}[Structural asymmetry of veto power]
\label{prop:veto-asymmetry}
The logical conditions governing division and cohesion are
asymmetric. Collapse requires only unilateral condition
failure, governed by a logical disjunction ($a(i,g)=1 \lor a(j,g)=1$).
Confirmation of cohesion requires a bilateral match, governed by a
logical conjunction ($a(i,g)=1 \land a(j,g)=1$).
\end{proposition}

\begin{proof}
Immediate from Definition~\ref{def:axis-satisfaction} (whose
antecedent is the disjunction $a(i,g)=1 \lor a(j,g)=1$, inherited by
the maintenance predicate of
Definition~\ref{def:post-event-maintenance}) and
Definition~\ref{def:confirmed-shared-condition}
(whose confirmation predicate involves the conjunction $a(i,g)=1
\land a(j,g)=1$).
\end{proof}

\begin{remark}[Structural basis of pairwise stability]
\label{rem:pairwise-stability}
In strategic network formation
\citep{jackson1996strategic}, the rule that link severance is
unilateral while link formation is bilateral is introduced as a
behavioral premise about consent among utility-maximizing players.
Proposition~\ref{prop:veto-asymmetry} recovers this asymmetry as a
logical property of constraint satisfaction: collapse is triggered
by the disjunction of withdrawal conditions, whereas a confirmed
shared condition requires their conjunction. This is a structural
analogue of the consent asymmetry, not a recovery of pairwise
stability as an equilibrium concept: utilities and link-formation
choices are absent from the model.
\end{remark}

\subsection{Impossibility of No-Collapse without Universal Compatibility}
\label{subsec:impossibility}

\begin{definition}[Universally compatible positioning]
\label{def:universally-compatible-positioning}
A positioning event $E : \mathcal{P} \to \mathcal{L}$ is
\emph{universally compatible} with respect to $g$ if
$f_g(E(i), E(j)) = 1$ for all $\{i,j\} \in \binom{\mathcal{P}}{2}$.
Equivalently, the pullback compatibility graph on the player
set---vertices $\mathcal{P}$, an edge $\{i,j\}$ iff
$f_g(E(i),E(j)) = 1$---is complete. When $E$ is non-injective, this
requires $f_g(\ell, \ell) = 1$ at every position $\ell$ occupied by
more than one player; completeness of the image $E(\mathcal{P})$ as
a subgraph of the compatibility graph on $\mathcal{L}$ does not by
itself test these diagonal values.
\end{definition}

Universal compatibility differs from degeneracy (Case~1 of
Theorem~\ref{thm:relativity}): degeneracy is relative to a specific
$\mathcal{R}_{\textnormal{initial}}$, whereas universal
compatibility is a design-level guarantee, forcing $\phi \equiv 1$
regardless of the initial relations; under the equivalence-relation
specialization it amounts to class-level homogenization,
$E(\mathcal{P})$ lying in a single class of $f_g$.

\begin{theorem}[Impossibility of no-collapse without universal compatibility]
\label{thm:impossibility}
Let $E : \mathcal{P} \to \mathcal{L}$ be a positioning event and $g$ a
position-dependent gain axis, with the positional structure
$(\mathcal{P}, \mathcal{L}, f_g, E)$ held fixed. Then no $g$-induced
fragmentation occurs---$\operatorname{Frag}_g(E)$ fails
(Definition~\ref{def:g-fragmentation})---for every relational layer
containing $g$ and sharing this positional structure, i.e., for
every admissible minimum-condition profile $a$ and every admissible
initial relation set $\mathcal{R}_{\textnormal{initial}}$ on such a
layer (Definition~\ref{def:initial-relation-set}), if and only if
$E$ is universally compatible with respect to $g$
(Definition~\ref{def:universally-compatible-positioning}), i.e.,
$f_g(E(i), E(j)) = 1$ for all $\{i,j\} \in \binom{\mathcal{P}}{2}$.
\end{theorem}

\begin{proof}
$(\Leftarrow)$ If $E$ is universally compatible, then $f_g(E(i),
E(j)) = 1$ for every pair $\{i,j\} \in \binom{\mathcal{P}}{2}$.
Hence for any
$R_{ij} \in \mathcal{R}_{\textnormal{initial}}$ with $g \in G_R$,
the consequent of Definition~\ref{def:axis-satisfaction} holds, so
$R_{ij} \models_E g$; thus no relation collapses due to $g$ and
$\operatorname{Frag}_g(E)$ fails.
$(\Rightarrow)$ If $E$ is not universally compatible, there exists
$\{i,j\} \in \binom{\mathcal{P}}{2}$ with $f_g(E(i), E(j)) = 0$.
Choose an admissible minimum-condition profile with
$a(i, g) = a(j, g) = 1$ and an admissible
$\mathcal{R}_{\textnormal{initial}} \ni R_{ij}$ on the resulting
layer; then $\{i,j\} \in \mathcal{E}_g^-$ and
$R_{ij} \not\models_E g$, so $\operatorname{Frag}_g(E)$ holds.
\end{proof}

For a fixed layer and initial set with
$\mathcal{E}_g^- \neq \emptyset$, the theorem reduces to a
per-instance dichotomy: $g$ induces no collapse among the relations
on $\mathcal{E}_g^-$ iff $\phi = f_g(E(\cdot),E(\cdot)) \equiv 1$
there, and any $0$ of $\phi$ on $\mathcal{E}_g^-$ forces at least
one collapse (Definition~\ref{def:axis-satisfaction}, whose
antecedent holds non-vacuously on $\mathcal{E}_g^-$).

Hence $f_g \equiv 1$ (existence-dependence) is the unique dependency
type guaranteeing the absence of $g$-induced collapse uniformly over
all admissible triples $(E, a, \mathcal{R}_{\textnormal{initial}})$.

\subsection{Structural Inescapability of Incompatible Dyads}
\label{subsec:outgroup}

\begin{corollary}[Inescapability of incompatible dyads]
\label{cor:outgroup-inescapability}
Let $g$ be position-dependent and non-degenerate under $E$
(Definition~\ref{def:g-relevant-degeneracy}). For any pair $\{u,v\}$ for which $g$ is a confirmed shared
condition under $E$, there exists at least one pair $\{i,j\} \in
\mathcal{E}_g^-$ such that $f_g(E(i), E(j)) = 0$ and $R_{ij}$
collapses.
\end{corollary}

\begin{proof}
If $\{u,v\}$ attains a confirmed shared condition, $\{u,v\} \in
\mathcal{E}_g^-$ and $f_g(E(u),E(v))=1$. Non-degeneracy makes
$\phi$ non-constant on $\mathcal{E}_g^-$, so some $\{i,j\} \in
\mathcal{E}_g^-$ has $f_g(E(i),E(j))=0$; by
Theorem~\ref{thm:relativity}, $R_{ij}$ collapses.
\end{proof}

\begin{corollary}[Partition of positions under equivalence-relation compatibility]
\label{cor:outgroup-partition-equivalence-case}
If $f_g$ defines an equivalence relation on $\mathcal{L}$
(Proposition~\ref{prop:exclusive-regions}), then for every $\{i,j\}
\in \mathcal{E}_g^-$ with $f_g(E(i),E(j))=0$, $E(i)$ and $E(j)$ lie
in distinct equivalence classes of $f_g$.
\end{corollary}

\begin{proof}
If $E(i)$ and $E(j)$ lay in the same equivalence
class, then $f_g(E(i),E(j))=1$ by definition of the equivalence
classes of $f_g$, contradicting the hypothesis. Hence they lie in
distinct classes.
(When $g$ is non-degenerate under $E$, at least one such pair exists by
Corollary~\ref{cor:outgroup-inescapability}.)
\end{proof}

Under this specialization the incompatible dyads of
Corollary~\ref{cor:outgroup-inescapability} become inter-class
pairs; the sociological reading as \emph{out-group pairs} is exact.

\section{Measure-Level Results}
\label{sec:measure}

\begin{convention}[Complete-dyad single-axis specialization]
\label{conv:complete-dyad}
Throughout Section~\ref{sec:measure} we specialize the initial
relational layer to the \emph{complete-dyad single-axis} setting:
$\mathcal{R}_{\textnormal{initial}}^{g}
    := \{R_{ij}^{g} : \{i,j\} \in \binom{\mathcal{P}}{2}\}$,
with $g$ the unique active position-dependent gain axis governing
maintenance and with the requirement bilateral throughout:
$a(i,g) = 1$ for every $i \in \mathcal{P}$. Under this
specialization,
$\mathcal{E}^{-} = \mathcal{E}_g^{-} = \binom{\mathcal{P}}{2}$,
every maintained edge carries a confirmed shared condition
(Definition~\ref{def:confirmed-shared-condition}), and for every $t$ the
threshold graph $\mathcal{G}_t = (\mathcal{P}, \mathcal{E}_t)$
(Definition~\ref{def:threshold-graph}) is exactly the instance of
the maintained-relation graph obtained by
evaluating maintenance with $f_t$ in place of $f_g$
(Definition~\ref{def:post-event-maintenance}).
Section~\ref{sec:existence}'s results continue to hold under this
specialization.
\end{convention}

\begin{remark}[Scope of the complete-dyad specialization]
\label{rem:canonical-specialization}
The specialization isolates the contribution of the compatibility
function: with every pair carrying an initial relation, all
variation in the maintained network is attributable to $f_t$ alone.
The assumption is substantive: a partial initial layer preserves the
restriction structure $\mathcal{E}_t \cap \mathcal{E}^-$ and the
local transition algebra (Lemma~\ref{lem:transition-law}), but not
the normalized interpretation or the complete-path theorems; see
Appendix~\ref{app:independence}.
\end{remark}

\subsection{Three Quantities and the Decomposition Identity}
\label{subsec:decomposition}

Recall $N = |\mathcal{P}|$ (Definition~\ref{def:player-set}). Given a relaxation family $\{f_t\}$
(Definition~\ref{def:relaxation-family}), let $M(\mathcal{G}_t)$
and $Q(\mathcal{G}_t)$ denote the edge count and intra-component
pair count of $\mathcal{G}_t$
(Definition~\ref{def:graph-invariants}). The survival rate
$M(\mathcal{G}_t) / \binom{N}{2}$ is the fraction of initial dyadic
relations in $\mathcal{R}_{\textnormal{initial}}^{g}$ maintained by
$E$ at laxity $t$.

\begin{definition}[Survival, integration, cohesion]
\label{def:sigma-I-K}
For $t$ with $\mathcal{E}_t \neq \emptyset$:
\begin{itemize}
    \item \emph{Survival rate}
    $\sigma(t) := M(\mathcal{G}_t) / \binom{N}{2}$: equivalently
    $1 - (\text{fragmentation rate})$, cf.\ Definition~\ref{def:fragmentation}.
    \item \emph{Integration}
    $I(t) := Q(\mathcal{G}_t) / \binom{N}{2}$: the fraction of pairs
    mutually reachable in $\mathcal{G}_t$.
    \item \emph{Cohesion} (intra-component density)
    $K(t) := M(\mathcal{G}_t) / Q(\mathcal{G}_t)$: the density of
    maintained relations among mutually reachable pairs, defined only
    when $\mathcal{E}_t \neq \emptyset$.
\end{itemize}
$\sigma$, $I$, $K$ are constant between transitions
(Definition~\ref{def:threshold-transition}).
\end{definition}

Fragmentation of Definition~\ref{def:fragmentation} corresponds to
$1 - \sigma$, and $K = 1$ (every mutually reachable pair
maintained) is substantially stronger than the existence-level
occurrence of local cohesion
(Remark~\ref{rem:three-tier-cohesion}). $K = M/Q$ is
\emph{intra-component edge density}, distinct from the
node-connectivity structural cohesion of \citet{moody2003structural}
(Section~\ref{subsec:feasibility}); the term ``cohesion'' is
retained because $K$ is the pair-density factor of the
decomposition.

\begin{proposition}[Decomposition identity]
\label{prop:decomposition}
Whenever $K$ is defined, $\sigma(t) = I(t) \cdot K(t)$.
\end{proposition}

\begin{proof}
Immediate: $M / \binom{N}{2} = (Q / \binom{N}{2}) \cdot (M / Q)$.
\end{proof}

\begin{remark}[$K = 1$ and Davis-clusterable structure]
\label{rem:davis-factions}
$K(t) = 1$ iff every component of $\mathcal{G}_t$ is complete:
$\mathcal{G}_t$ is a disjoint union of cliques. Under the canonical
signed completion (positive within components, negative between; an
auxiliary signed extension used only for the Davis correspondence),
such a network instantiates a Davis-clusterable configuration in the
sense of
\citep{davis1967clustering}: the components are the factions of
Davis's structure theorem.
\end{remark}

\begin{lemma}[Monotone components]
\label{lem:sigma-I-monotone}
$\sigma$ and $I$ are nondecreasing in $t$.
\end{lemma}

\begin{proof}
$\mathcal{E}_t \subseteq \mathcal{E}_{t'}$ for $t \le t'$ gives $M$
nondecreasing. Merging components of sizes $n_1, n_2$ increases $Q$
by $n_1 n_2 > 0$.
\end{proof}

\subsection{The Local Law: Marginal Fill Test}
\label{subsec:local-law}

Write $K^-$ ($K^+$) for cohesion just before (after) a transition.

\begin{lemma}[General transition law]
\label{lem:transition-law}
Consider any transition at which $K^-$ is defined, with changes
$\Delta M \ge 1$ and $\Delta Q \ge 0$ (ties permitted).
\begin{enumerate}
    \item If $\Delta Q = 0$, then $K^+ > K^-$.
    \item If $\Delta Q > 0$, then $K^+ \gtreqless K^-$ iff
    $\Delta M / \Delta Q \gtreqless K^-$.
\end{enumerate}
\end{lemma}

\begin{proof}
From
$K^+ - K^-
= (Q \cdot \Delta M - M \cdot \Delta Q) / (Q\,(Q + \Delta Q))$
with positive denominator.
\end{proof}

The ratio $\Delta M / \Delta Q$ is the \emph{marginal fill ratio}:
cohesion decreases at a transition precisely when this ratio falls
below $K^-$.

\begin{definition}[Cohesion-synchronized transition]
\label{def:synchronised}
A transition at which $K^-$ is defined is
\emph{cohesion-synchronized} if $\Delta Q = 0$ or
$\Delta M / \Delta Q \ge K^-$.
\end{definition}

By Lemma~\ref{lem:transition-law}, a transition is synchronized iff
$K$ does not decrease across it. The simple-transition specialization ($\Delta M = 1$) is developed
in Appendix~\ref{app:measure-details}; it covers merger dilution
when $n_1 n_2 > 1/K^-$ and its consequences when $K^- = 1$.

\subsection{Universality: Every Complete Strict Path is Non-Monotone}
\label{subsec:universality}

\begin{definition}[Complete strict relaxation path]
\label{def:complete-strict-path}
A relaxation family (Definition~\ref{def:relaxation-family}) realizes
a \emph{complete strict path} if its realized chain of edge sets runs
$\mathcal{E}^{(1)} \subsetneq \cdots \subsetneq \mathcal{E}^{(m)}
    = \binom{\mathcal{P}}{2}$
with $|\mathcal{E}^{(1)}| = 1$ and every transition simple.
\end{definition}

\begin{theorem}[Non-monotonicity of cohesion]
\label{thm:universal-non-monotone}
Let $N \ge 3$. Along every complete strict path, $K$ is non-monotone:
$K = 1$ at the first step, $K = 1$ at the last, and there is an
intermediate step with $K < 1$, while $I$ rises from
$1/\binom{N}{2}$ to $1$.
\end{theorem}

\begin{proof}
At the first step $M = Q = 1$, so $K = 1$; at the last step
$\mathcal{G}_m = K_N$, so $M = Q = \binom{N}{2}$ and $K = 1$. (For
illustration, the minimal case $N = 3$ with distinct distances gives
the trajectory
$(M, Q, K) = (1,1,1) \to (2,3,2/3) \to (3,3,1)$.)

Let step $\tau$ be the first at which the threshold graph is
connected. By minimality of $\tau$, $\mathcal{G}_\tau$ is not the
complete graph: if it were, its predecessor (differing by exactly
one edge) would be $K_N$ minus one edge, which is connected for
$N \ge 3$ and would already have witnessed connectivity at step
$\tau - 1$. Hence $\mathcal{G}_\tau$ is connected and incomplete:
$I = 1$ and $K = \sigma < 1$ at step $\tau$. A monotone
nondecreasing $K$ would force $K(\tau) \ge 1$; a monotone
nonincreasing $K$ would force the final value $\le K(\tau) < 1$.
Both fail.
\end{proof}

\subsection{Structural Crossing under Scale Separation}
\label{subsec:crossing}

\begin{definition}[$(s, S, D)$-clustered configuration]
\label{def:clustered-config}
Positions $E(\mathcal{P})$ admit a partition into clusters
$B_1, \dots, B_k$, $k \ge 2$, each of size $n \ge 2$ ($N = kn$), and
parameters $s < S \le D$ such that:
\begin{enumerate}
    \item[(i)] $s = \max\{ \mathrm{dist}(\ell, \ell')
    : \ell, \ell' \text{ in the same cluster}\}$ (attained);
    \item[(ii)] $\mathrm{dist}(\ell, \ell') \ge S$ for positions in
    different clusters;
    \item[(iii)] the clusters can be ordered so that consecutive
    clusters contain a pair at distance $\le D$, and every pair in
    non-consecutive clusters is at distance $> D$.
\end{enumerate}
\end{definition}

The trajectories of $I$ and $K$ are step functions; by a
\emph{crossing} we mean the reversal of their order along the
family---equality need not be attained at any parameter value.

\begin{theorem}[Structural crossing]
\label{thm:structural-crossing}
For an $(s, S, D)$-clustered configuration under the
distance-threshold family:
\begin{enumerate}
    \item For every $d \in [s, S)$: $K(d) = 1$ and
    $I(d) = (n-1)/(kn-1) < 1/k$.
    \item At $d = D$: $I(D) = 1$ and
    $K(D) \le [(n-1) + 2n(k-1)/k] / (kn-1)$, with sharp asymptotic
    value $(3k - 2)/k^2$ under complete cross-connection at scale
    $D$ (Proposition~\ref{prop:envelope}); for $k \ge 3$ the
    asymptotic value bounds $K(D)$ uniformly over all cluster sizes
    $n$ (Remark~\ref{rem:uniform-envelope}).
\end{enumerate}
As $d$ increases from any $d_1 \in [s, S)$ to $D$, integration
rises to $1$ and cohesion drops from $1$ to $K(D)$.

\emph{For $k \ge 3$}: $I(D) > K(D)$ strictly. For $k \ge 5$ the
margin is uniform, $\min(I(D) - I(d_1),\, K(d_1) - K(D)) > 1 - 4/k$.

\emph{For $k = 2$}: $I(D) \ge K(D)$, with equality attained when
the two clusters are completely cross-connected at scale $D$.
\end{theorem}

\begin{proof}
Deferred to Appendix~\ref{app:measure-details}.
\end{proof}

The separation $s < S$ creates the nonempty threshold interval on
which clusters are internally complete but mutually disconnected;
the depth of the crossing is controlled primarily by the cluster
number $k$, with the ratios $S/s$ and $D/S$ absent from the depth
bound (envelope derived in Appendix~\ref{app:measure-details}).

\subsection{Characterization of Monotone Trajectories}
\label{subsec:characterization}

Theorem~\ref{thm:synchronisation} characterizes monotone
trajectories exactly;
Propositions~\ref{prop:immediate-connectivity}
and~\ref{prop:ultrametric-escape} then exhibit two global regimes
certifying its condition.

\begin{theorem}[Synchronization characterization]
\label{thm:synchronisation}
Consider a finite relaxation trajectory from the empty graph to the
complete graph on $N \ge 2$ players, and evaluate $K$ from the first
nonempty edge set onward. Then $K$ is monotone if and only if every
transition after the first nonempty one is cohesion-synchronized
(Definition~\ref{def:synchronised}).
\end{theorem}

\begin{proof}
On a complete trajectory $K = 1$ at the endpoint and $K \le 1$
throughout. If $K$ is nonincreasing, $K \ge 1$ everywhere, so
$K \equiv 1$. Hence monotone $\iff$ nondecreasing on complete
trajectories. $K$ is constant between transitions, and by
Lemma~\ref{lem:transition-law} nondecreasing across a transition iff
that transition is synchronized. (If the first nonempty edge set is
already complete, no subsequent transitions exist and $K \equiv 1$.)
\end{proof}

\begin{corollary}[Global necessity of ties]
\label{cor:ties-necessary}
Let $N \ge 3$. Any relaxation trajectory from the empty graph to the
complete graph along which $K$ is monotone must contain at least one
non-simple transition.
\end{corollary}

\begin{proof}
If the first nonempty transition introduces two or more edges, it is
non-simple. If it introduces exactly one edge and all subsequent
transitions are simple, the trajectory is a complete strict path;
Theorem~\ref{thm:universal-non-monotone} yields non-monotonicity.
\end{proof}

\begin{proposition}[Immediate connectivity: $\Delta Q$ frozen]
\label{prop:immediate-connectivity}
Suppose the first nonempty realized edge set $\mathcal{E}^{(1)}$
yields a connected graph. Then $I = 1$ from that point on, every
subsequent transition has $\Delta Q = 0$, hence is synchronized, and
$K = \sigma$ is nondecreasing on its whole domain.
\end{proposition}

\begin{proof}
Connectivity is preserved under edge addition, so $Q =
\binom{N}{2}$ and $I \equiv 1$. Lemma~\ref{lem:transition-law}(1)
applies; $K = \sigma$ is nondecreasing by
Lemma~\ref{lem:sigma-I-monotone}.
\end{proof}

\begin{proposition}[Ultrametric hierarchy: $\Delta M = \Delta Q$]
\label{prop:ultrametric-escape}
Let $\mathrm{dist}$ be the metric induced on the finite position set
$E(\mathcal{P})$.
\begin{enumerate}
    \item[\textnormal{(a)}] \textbf{Forward.} If $\mathrm{dist}$ is
    an ultrametric, then every threshold graph $\mathcal{G}_r$ is a
    disjoint union of cliques; consequently $K(r) = 1$ whenever
    $\mathcal{G}_r$ is nonempty, and at every transition $\Delta M =
    \Delta Q$ (the synchronization condition holds with equality).
    \item[\textnormal{(b)}] \textbf{Converse.} If every threshold
    graph $\mathcal{G}_r$ is a disjoint union of cliques, then the
    induced finite metric satisfies the ultrametric inequality.
\end{enumerate}
\end{proposition}

\begin{proof}
Deferred to Appendix~\ref{app:measure-details}. The correspondence
between dendrograms and finite ultrametric spaces underlying the
statement is classical
\citep{johnson1967hierarchical, carlsson2010clustering}; the present
statement translates it into the threshold-graph framework and is not
claimed as new.
\end{proof}

Both regimes are global certificates for the same local
synchronization condition, and both supply the ties required by
Corollary~\ref{cor:ties-necessary}: an ultrametric forces isosceles
triples with the two largest distances equal, and immediate
connectivity on $N \ge 3$ players requires $\mathcal{E}^{(1)}$ to
contain at least $N-1$ edges.

Euclidean realizability restrictions on the two escape regimes are
developed in Appendix~\ref{app:euclidean-restrictions}.

\section{Discussion}
\label{sec:discussion}

\subsection{Scope and Extensions}
\label{subsec:scope}

Three scope choices frame the results. The comparison between pre-
and post-event states is logical rather than temporal: parameter
values in the relaxation family
(Definition~\ref{def:relaxation-family}) index configurations, not
time, and the results characterize what states the compatibility
function admits, not the paths by which players arrive at them. The
positioning event $E$ is exogenous
(Remark~\ref{rem:scope-exogeneity}). The algebraic statements of
Section~\ref{sec:measure} are proved for a single position-dependent
gain axis $g$; the framework extends to multiple axes by intersection
over the relational layer (Postulate~\ref{po:layer-fixation}), but
the sharpness of the decomposition $\sigma = I \cdot K$ and the
universal non-monotonicity of $K$ depend on operating within one
axis at a time. A systematic audit of the assumptions and their
scope boundaries is given in Appendix~\ref{app:independence}.

\paragraph{Dynamic extensions.}
The feasibility layer bounds the space within which any dynamic
extension of the framework must unfold. Two extension classes are
immediate. In \emph{position-drift dynamics}, players update $E(i)$
over time under strategic, imitative, or exogenous pressure, so
Theorem~\ref{thm:relativity} constrains the state space of that
trajectory rather than an equilibrium. In \emph{axis-set dynamics},
the set of active gain axes itself evolves, shifting the
identity of the in-group/out-group partition rather than only its
intensity. In both cases the
decomposition $\sigma = I \cdot K$ applies snapshot by snapshot. The
non-monotonicity theorem constrains a dynamic extension only when
the induced sequence of maintained networks is itself a complete
strict relaxation path
(Definition~\ref{def:complete-strict-path}); general position or
axis dynamics, in which edges may disappear and reappear, need not
satisfy this condition.

\subsection{Empirical Interpretation and Identification}
\label{subsec:misattribution-implications}

Behavioral, affective, and strategic accounts of division and
cohesion attribute observed patterns to a proximate cause operating
one layer above the feasibility constraint.
Corollary~\ref{cor:outgroup-inescapability} makes the division of
labor concrete: an
affective account can explain which severed pair intensifies into
hostility, and how it is emotionally sustained, but not whether some
severed pair must exist.

\paragraph{Empirical handles.}
The decomposition $\sigma = I \cdot K$
(Proposition~\ref{prop:decomposition}) and the universal
non-monotonicity of $K$
(Theorem~\ref{thm:universal-non-monotone}) yield quantities directly
computable from a graph paired with a position map. Panel settings
in which players carry a positional coordinate---roll-call votes,
co-sponsorship dyads with ideology scores, occupational categories
with mobility ties---admit direct computation of $\sigma$, $I$, and
$K$ at any indexed configuration. The cohesion quantities of
\citet{moody2003structural} and the phase-transition
indicators of \citet{pham2020effect} sit above this
decomposition and are complementary to it.

Testing the theory requires an \emph{external} indexing of the
compatibility function: the identity $\sigma = I \cdot K$ and the
transition algebra (Lemma~\ref{lem:transition-law}) are arithmetic
and cannot fail on any observed graph. Under the distance-threshold
specialization (Definition~\ref{def:metric-specialization}), the
threshold $d$ is specified independently of the observed relational
outcomes, and
Theorem~\ref{thm:universal-non-monotone} predicts non-monotone $K(d)$
along any complete strict path through $E(\mathcal{P})$. With positions and
edges independently measured, observation of a monotone
$K(d)$-trajectory along such an ordering constitutes a specification
diagnostic against the distance-threshold specialization. When the positional data admit an
$(s, S, D)$-clustered decomposition,
Theorem~\ref{thm:structural-crossing} sharpens the prediction to
the joint trajectory $(I(d), K(d))$.

\citet{lorenz2026diversity} find that the structural cohesion of
adolescent friendship networks is lowest at intermediate ethnic
diversity and recovers at both low and high diversity---a non-monotone
cohesion profile in a positional-compatibility covariate, matching the
shape predicted by Theorem~\ref{thm:universal-non-monotone}. The
reported associations are small, and diversity there indexes
configurations cross-sectionally rather than tracing a complete strict
relaxation path, so the agreement is one of shape rather than a direct
test.

The primitives $(G, a, f_g)$ are not identified by $\sigma$, $I$, $K$
alone. Empirical content is comparative: it requires an independent
handle on $f_g$ (an external threshold parameter, a declared
withdrawal condition, or a natural experiment on $E$). This
requirement is not optional: for an injective position map, any
graph on $\mathcal{P}$ can be represented ex post by a suitably
chosen compatibility function, so $f_g$ must be fixed independently
of the network it is intended to explain.

\printbibliography

\clearpage
\appendix
\renewcommand{\thetheorem}{\Alph{section}.\arabic{theorem}}

\section{Assumption Audit and Scope Boundaries}
\label{app:independence}

For each core assumption of Theorems~\ref{thm:relativity}
and~\ref{thm:impossibility}, we identify its role and note when its
removal changes the theorems' scope. Assumptions are classified along
five axes: \emph{constitutive} (defining the object of study),
\emph{logical} (invoked by the proofs), \emph{non-degeneracy}
(distinguishing structural from vacuous cases),
\emph{specialization} (activating
Section~\ref{sec:measure}), and \emph{interpretive} (licensing
sociological readings). The scope-boundary witnesses below use the
minimal setting $\mathcal{P}=\{1,2,3\}$, $|\mathcal{L}|\ge 2$, and
$\mathcal{R}_{\textnormal{initial}} \ne \emptyset$; the theorems
themselves remain stated in full generality.

\paragraph{Constitutive: pairwise reduction (Postulate~\ref{po:reduct-pairwise}).}
Fixes the object of study as dyadic relations. \emph{Boundary:} Under a triadic joint-feasibility predicate
$F_g(E(1),E(2),E(3))$, group-level satisfaction can absorb pairwise
incompatibilities. This changes the object of study;
complementary-necessity results for higher-order relations require
separate development.

\paragraph{Constitutive: fixed relational layer (Postulate~\ref{po:layer-fixation}).}
Fixes a single relational layer as the domain of
maintenance. \emph{Boundary (aggregation masking):} Aggregating a second
layer with $f_g \equiv 1$ into the union network means every pair
retains a connection at the aggregated level. Layer-1 fragmentation
still holds within Layer 1 but is masked once layers are merged.

\paragraph{Logical: minimum condition (Axiom~\ref{ax:minimum-condition}).}
Substantive premise behind
Definition~\ref{def:post-event-maintenance}: it justifies encoding
maintenance as the conjunctive satisfaction of every active gain
axis. \emph{Boundary:} Weakening the axiom (unsatisfied $g$ does
not force withdrawal) requires simultaneously weakening
Definition~\ref{def:post-event-maintenance}; the theorems then no
longer apply because their premise ``maintained after $E$'' carries a
different meaning.

\paragraph{Logical: single-valued positioning (Definition~\ref{ax:unique-position}).}
Invoked in the Fragmentation branch of
Theorem~\ref{thm:relativity}: fixed single-valued positioning makes
an incompatibility irresolvable. \emph{Boundary:} With set-valued positioning
$\tilde{E}:\mathcal{P}\to 2^{\mathcal{L}} \setminus \{\emptyset\}$,
an incompatible pair can be resolved by choosing a compatible
alternative position; the theorem's contradiction step fails.

\paragraph{Non-degeneracy: both compatible and incompatible relevant dyads.}
Case hypothesis of Theorem~\ref{thm:relativity}
(Case 2): $\phi$ is not constant on $\mathcal{E}_g^-$. Separates
the substantive complementary-necessity regime from the degenerate
one. \emph{Countermodel:} Let $\mathcal{L}=\{a,b\}$, $f_g$
the equality compatibility
($f_g(\ell,\ell')=\mathbf{1}\{\ell = \ell'\}$), and $E$ map every
relevant player to $a$. Here $g$ is position-dependent (not
existence-dependent, since $f_g(a,b)=0$), yet on $E(\mathcal{P})$
the map $\phi \equiv 1$: the system falls under Case 1 (degenerate),
so no fragmentation occurs. Case 2 does not apply.

\paragraph{Specialization: Section~\ref{sec:measure} setup (Convention~\ref{conv:complete-dyad}).}
Section~\ref{sec:measure}'s measure-level results are
stated under a complete-dyad single-axis specialization; this is a
scope declaration for Section~\ref{sec:measure}, not an assumption of the existence-level
theorems. \emph{Boundary:} Partial dyad coverage or multiple active
axes constitute a separate setting, not a reindexing: under
$|\mathcal{E}^-|$-normalization the integration term
$Q/|\mathcal{E}^-|$ can exceed~$1$ ($Q$ counts mutually reachable
pairs that need not be initial dyads), and the complete-path results
of Section~\ref{sec:measure} rely on the terminal graph being
complete. Only the arithmetic identity $M = Q \cdot (M/Q)$ and the
transition law (Lemma~\ref{lem:transition-law}) transfer unchanged
(Remark~\ref{rem:canonical-specialization}).

\paragraph{Interpretive: equivalence-relation structure of $f_g$.}
Neither Theorem~\ref{thm:relativity} nor
Theorem~\ref{thm:impossibility} assumes $f_g$ is transitive; both
hold for any symmetric $f_g$. Transitivity is invoked only by
Proposition~\ref{prop:exclusive-regions} and the equivalence-case
corollary (Corollary~\ref{cor:outgroup-partition-equivalence-case}). \emph{Boundary:} Sociological readings that presume a
genuine in-group/out-group partition are exact only under the
equivalence-relation specialization. Under a general symmetric
$f_g$, the theorems yield compatibility cliques and incompatible
dyads, not partitions.

\paragraph{Entailed: nontriviality of the position space ($|\mathcal{L}| \geq 2$).}
This is not an independent assumption: Non-Null Compatibility
(Assumption~\ref{ass:non-null}) together with position-dependence of
$g$ forces $|\mathcal{L}| \ge 2$, since a singleton position space
admits only the constant $f_g \equiv 1$.

\section{Measure-Level Details}
\label{app:measure-details}

\subsection{Simple-Transition Specialization of the Local Law}
\label{subsec:simple-transition}

Call a transition \emph{simple} if $\Delta M = 1$: exactly one pair
enters. In that case, the general transition law
(Lemma~\ref{lem:transition-law}) reduces to a merger-dilution
statement in which the geometry of the two components being joined
determines the outcome directly.

\begin{lemma}[Merger dilution, simple transitions]
\label{lem:merger-dilution}
Let the $r$-th transition be simple with new edge $e$, and assume
$\mathcal{E}_r^- \neq \emptyset$, so that $K^-$ is defined.
\begin{enumerate}
    \item If $e$ joins two nodes in the same component of
    $\mathcal{G}_r^-$ (\emph{densification}: $\Delta Q = 0$), then
    $K$ strictly increases.
    \item If $e$ joins components of sizes $n_1, n_2$
    (\emph{merger}: $\Delta Q = n_1 n_2$), then $K$ strictly
    decreases iff $n_1 n_2 > 1/K^-$; $K$ is unchanged iff
    $n_1 n_2 = 1/K^-$; otherwise $K$ strictly increases.
\end{enumerate}
At the very first edge of a trajectory $K^-$ is undefined; that step
initializes $K = 1$.
\end{lemma}

\begin{proof}
Special case of Lemma~\ref{lem:transition-law} with $\Delta M = 1$
and $\Delta Q \in \{0, n_1 n_2\}$.
\end{proof}

In particular, from a fully cohesive state ($K^- = 1$) any merger
not joining two singletons strictly decreases cohesion, while a
two-singleton merger ($n_1 n_2 = 1 = 1/K^-$) leaves $K = 1$.

\subsection{Proof of Theorem~\ref{thm:structural-crossing}}
\label{subsec:proof-crossing}

Recall the setup: an $(s, S, D)$-clustered configuration
(Definition~\ref{def:clustered-config}) partitions the positions
into $k \ge 2$ clusters of equal size $n \ge 2$ ($N = kn$), with
$s = \max\{ \mathrm{dist}(\ell, \ell') : \ell, \ell'
\text{ same-cluster}\}$ attained, $\mathrm{dist} \ge S$ across
clusters, and consecutive-cluster bridges at distance $\le D$ but
non-consecutive-cluster distances $> D$.

\begin{proof}[Proof of Theorem~\ref{thm:structural-crossing}]
\textbf{(1) Plateau interval $d \in [s, S)$.} By condition~(i) of
Definition~\ref{def:clustered-config}, every within-cluster pair
lies in $\mathcal{E}_d$; by condition~(ii), no between-cluster pair
does. Each component of $\mathcal{G}_d$ is therefore a complete
cluster of size $n$: $M = Q = k \binom{n}{2}$, whence $K(d) = 1$
and
\[
I(d) = k\binom{n}{2}\big/\binom{kn}{2}
= (n - 1)\big/(kn - 1).
\]
The bound $(n-1)/(kn-1) < 1/k$ is equivalent to
$k(n-1) < kn - 1$, i.e., $k > 1$.

\textbf{(2) Endpoint $d = D$.} At $d = D$: within-cluster pairs are
in $\mathcal{E}_D$ (since $s \le D$); by condition~(iii), each pair
of consecutive clusters is joined, so $\mathcal{G}_D$ is connected
and $I(D) = 1$. Between-cluster edges can only join consecutive
clusters (all other between-cluster distances exceed $D$), so
\[
M(\mathcal{G}_D) \le k \binom{n}{2} + (k-1) n^2.
\]
Since $I(D) = 1$ we have $K(D) = \sigma(D)
= M(\mathcal{G}_D) / \binom{kn}{2}$; substituting the bound gives
\[
K(D) \le \frac{k \binom{n}{2} + (k-1) n^2}{\binom{kn}{2}}
= \frac{(n-1) + 2n(k-1)/k}{kn - 1}.
\]
Bounding $(n - 1) < n$ and $2n(k-1)/k < 2n$ yields
$K(D) < 3n/(kn - 1)$, and $3n/(kn - 1) \le 4/k$ is equivalent to
$3nk \le 4kn - 4$, i.e., $4 \le kn$, which holds since
$k, n \ge 2$. (The preceding inequality is strict, so
$K(D) < 4/k$ also when $kn = 4$.)

\textbf{Crossing.} As $d$ increases from any $d_1 \in [s, S)$ to
$D$, $(I, K)$ moves from $(\alpha, 1)$ with $\alpha < 1/k$ to
$(1, K(D))$.

\emph{For $k \ge 3$}: condition~(iii) forces at least one pair of
non-consecutive clusters to remain at distance $> D$, so
$M(\mathcal{G}_D) < \binom{kn}{2}$ strictly, giving $K(D) < 1$.
The $k \ge 5$ margin follows from $K(D) < 4/k \le 1 - (1 - 4/k)$
and $I(D) - I(d_1) > 1 - 1/k \ge 1 - 4/k$.

\emph{For $k = 2$}: condition~(iii) is vacuously satisfied; the
$M$-bound above is achieved with equality when the two clusters
are completely cross-connected at scale $D$, giving
$M(\mathcal{G}_D) = \binom{2n}{2}$ and $K(D) = I(D) = 1$.
\end{proof}

\subsection{Envelope of the Crossing under Complete
Cross-Connection}
\label{subsec:envelope}

Hypothesis~(iii) of Definition~\ref{def:clustered-config} requires
only one bridging pair per consecutive-cluster gap, so $K(D)$ ranges
over an envelope determined by how many between-cluster edges are
actually present at scale $D$.

\begin{proposition}[Envelope range]
\label{prop:envelope}
Fix $k$ and let $n \to \infty$.
\begin{enumerate}
    \item \textbf{Bounded bridging.} If each consecutive-cluster
    gap is bridged by a uniformly bounded number of edges, then
    $M(\mathcal{G}_D) = k \binom{n}{2} + O(k)$ and
    $K(D) \to 1/k$.
    \item \textbf{Complete cross-connection.} If every consecutive
    pair of clusters is completely cross-connected at scale $D$,
    then $M(\mathcal{G}_D) = k \binom{n}{2} + (k-1) n^2$ and
    \[
    K(D) \to (3k - 2)/k^2 = 3/k - 2/k^2.
    \]
\end{enumerate}
\end{proposition}

\begin{proof}
Immediate from the expressions above and
$\binom{kn}{2} \sim (kn)^2/2$.
\end{proof}

\begin{remark}[Uniformity of the asymptotic bound]
\label{rem:uniform-envelope}
For $k \ge 3$ the finite-$n$ bound of
Theorem~\ref{thm:structural-crossing}(2) is strictly increasing in
$n$: writing $c := (3k-2)/k$, the bound equals $(cn - 1)/(kn - 1)$,
and the sign of its derivative in $n$ is that of
$k - c = (k-1)(k-2)/k > 0$. Hence $K(D) < (3k-2)/k^2$ for every
finite $n$, and the asymptotic value of
Proposition~\ref{prop:envelope} is simultaneously the supremum of
the finite-$n$ bounds. For $k = 2$ the bound equals $1$
identically, consistently with the non-strict crossing of
Theorem~\ref{thm:structural-crossing}.
\end{remark}

Complete cross-connection is geometrically realizable in one
dimension, so the upper envelope is attained. Place each $B_j$ inside an interval of length $s$, with
two of the $n$ points at the interval endpoints so that the diameter
$s$ is attained as required by
Definition~\ref{def:clustered-config}(i); space consecutive left
endpoints by $S + s$, and set $D = S + 2s$. Then within-cluster
distances are $\le s$; consecutive-cluster distances lie in
$[S, S + 2s] = [S, D]$, so all $n^2$ consecutive pairs are bridged;
and non-consecutive clusters are at distance $\ge 2S + s > D$
whenever $S > s$.

\subsection{Genericity of the Complete-Strict-Path Assumption}
\label{subsec:genericity}

\begin{proposition}[Genericity]
\label{prop:generic-strict}
Suppose the joint distribution of $(E(1), \dots, E(N))$ is
absolutely continuous with respect to Lebesgue measure on
$(\mathbb{R}^d)^N$. Then, outside a null set of position assignments,
all pairwise distances are distinct; in particular, for $N \ge 3$,
the distance-threshold family
(Definition~\ref{def:metric-specialization}) is almost surely a
complete strict path (Definition~\ref{def:complete-strict-path}),
and Theorem~\ref{thm:universal-non-monotone} applies.
\end{proposition}

\begin{proof}
For any two distinct unordered pairs
$\{i,j\} \neq \{k, \ell\}$ (sharing a vertex or not), the polynomial
$\|x_i - x_j\|^2 - \|x_k - x_\ell\|^2$ is not identically zero, so
its zero set is an algebraic hypersurface of Lebesgue measure zero
in $(\mathbb{R}^d)^N$. The event of any distance tie is contained
in the finite union of these hypersurfaces, hence has measure zero.
Outside the resulting null set, all $\binom{N}{2}$ distances are
distinct; the distance-threshold family then admits exactly
$\binom{N}{2}$ transitions, each adding exactly one edge, and
terminating in the complete graph.
\end{proof}

\subsection{Proof of Proposition~\ref{prop:ultrametric-escape}}
\label{subsec:proof-ultrametric}

\begin{proof}[Proof of Proposition~\ref{prop:ultrametric-escape}]
\sloppy
$(\Rightarrow)$ Assume $\mathrm{dist}$ is an ultrametric. If
$\mathrm{dist}(x, y) \le r$ and $\mathrm{dist}(y, z) \le r$, then
$\mathrm{dist}(x, z) \le \max\{\mathrm{dist}(x, y),
\mathrm{dist}(y, z)\} \le r$: adjacency at threshold $r$ is
transitive, so components of $\mathcal{G}_r$ are cliques and $M =
Q$, giving $K = 1$ at every nonempty threshold (in particular the
first). At any subsequent transition, if $x$ and $y$ newly share a
component, a path of edges of length $\le r$ joins them; iterating
the ultrametric inequality along this path gives $\mathrm{dist}(x,
y) \le r$, so every newly intra-component pair is itself an edge
of $\mathcal{G}_r$. Hence $\Delta M = \Delta Q$, and $K^- = 1$
gives equality in the marginal fill test.

$(\Leftarrow)$ Suppose every threshold graph $\mathcal{G}_r$ is a
disjoint union of cliques. For arbitrary $x, y, z$ set
$r = \max\{\mathrm{dist}(x, y), \mathrm{dist}(y, z)\}$ (the
threshold graph is defined at every real $r$, so this choice is
well-formed even if $r$ is not a realized transition value). Then
the edges $\{x, y\}$ and $\{y, z\}$ are in $\mathcal{G}_r$, so $x$
and $z$ lie in a common component; that component is complete by
hypothesis, so $\{x, z\}$ is an edge of $\mathcal{G}_r$, i.e.,
$\mathrm{dist}(x, z) \le r = \max\{\mathrm{dist}(x, y),
\mathrm{dist}(y, z)\}$.
\end{proof}

\section{Euclidean Restrictions on Escape Regimes}
\label{app:euclidean-restrictions}

The two escape regimes of Section~\ref{subsec:characterization} are
both non-generic under absolutely continuous Euclidean models
(Proposition~\ref{prop:generic-strict}), but their Euclidean
realizability differs. Immediate connectivity is readily realizable
in low dimension: equally spaced points on a line ($E(i) = i\delta$)
produce, at threshold $d = \delta$, all nearest-neighbor edges in a
single transition---a connected path, so by
Proposition~\ref{prop:immediate-connectivity} $I = 1$ from that step
onward and $K = \sigma$ rises monotonically to~$1$; any finite
rectangular subset of the integer lattice $\mathbb{Z}^2$ behaves
identically at $d = 1$. Such configurations lie in the null set of
Proposition~\ref{prop:generic-strict}. Ultrametric escape, by
contrast, is only partially realizable in low Euclidean dimension.

\begin{proposition}[Non-realizability of large equilateral
hierarchies]
\label{prop:no-large-equilateral}
The equilateral metric on $d + 2$ points admits no isometric
realization in $\mathbb{R}^d$.
\end{proposition}

\begin{proof}
An isometric realization of the equilateral metric on $m$ points
requires those points to be pairwise equidistant, hence to span an
affine simplex of dimension $m - 1$. This forces the ambient space
to have dimension at least $m - 1$, so no realization of the
equilateral metric on $m = d + 2$ points exists in $\mathbb{R}^d$.
\end{proof}

Since an equilateral metric on $m$ points is a finite ultrametric,
no fixed Euclidean dimension realizes all finite ultrametrics; an
equilateral triangle in $\mathbb{R}^2$ nevertheless realizes the
ultrametric on three points, giving a genuine ultrametric-escape
configuration in the plane.

\section{Notation}

\begin{longtable}{@{}p{0.28\textwidth} p{0.66\textwidth}@{}}
\caption{Notation used throughout the paper.}\label{tab:notation} \\
\toprule
\textbf{Symbol} & \textbf{Meaning} \\
\midrule
\endfirsthead

\multicolumn{2}{@{}l}{\textit{Table \ref{tab:notation} (continued).}} \\
\toprule
\textbf{Symbol} & \textbf{Meaning} \\
\midrule
\endhead

\midrule
\multicolumn{2}{r@{}}{\textit{Continued on next page.}} \\
\endfoot

\bottomrule
\endlastfoot

\( a(i,g) \) & Minimum-condition indicator: \( a(i,g)=1 \) if \( g \) is required by player \( i \) \\
\( f_g \) & Compatibility function associated with gain axis \( g \) \\
\( \mathcal{R}_{\textnormal{initial}}^{g} \) & Complete-dyad initial layer for a single axis \( g \) (\S\ref{sec:measure} specialization, Convention~\ref{conv:complete-dyad}) \\
\( \mathcal{E}_g^{-} \) & Set of \(g\)-shared-relevant initial edges \\

\midrule
\multicolumn{2}{@{}l}{\textit{Measure-level (Section~\ref{sec:measure}):}} \\
\midrule

\( \{f_t\}_{t \in T} \) & Relaxation family (nested compatibility functions) \\
\( \mathcal{G}_t \) & Threshold graph at parameter \( t \) \\
\( M(\mathcal{N}) \) & Edge count of a graph \( \mathcal{N} \) \\
\( Q(\mathcal{N}) \) & Intra-component pair count \( \sum_C \binom{|C|}{2} \) \\
\( \Delta M_r,\; \Delta Q_r \) & Increments in \( M,Q \) across the \(r\)-th transition \\
\( \sigma(t) \) & Survival rate \( M(\mathcal{G}_t)/\binom{N}{2} \) \\
\( I(t) \) & Integration \( Q(\mathcal{G}_t)/\binom{N}{2} \) \\
\( K(t) \) & Cohesion \( M(\mathcal{G}_t)/Q(\mathcal{G}_t) \) \\
\( K^-, K^+ \) & Cohesion values just before / after a transition \\
\( s, S, D \) & Cluster scale parameters (intra-cluster max, inter-cluster min, connection scale) \\
\( k, n \) & Cluster number and per-cluster size in \( (s,S,D) \)-clustered configurations \\
\end{longtable}

\end{document}